\documentclass[11pt]{article}

\usepackage{citesort}
\usepackage{graphicx}

\makeatletter
\@addtoreset{equation}{section} % reset eq. number to zero when
\renewcommand{\theequation}{\arabic{section}.\arabic{equation}}

\newcommand{\erf}{\mathop{\mathrm{erf}}}

\makeatother

\renewcommand{\r}{\mathbf{r}}
\newcommand{\Tr}{\mathop{\mathrm{Tr}}}

\begin{document}

\title{{\bf Pressures for a One-Component Plasma on a Pseudosphere}}

\author{{\bf R. Fantoni,}$^1$ {\bf B. Jancovici,}$^2$ {\bf and
G. T\'ellez}$^{2,3}$} 

\maketitle

\begin{abstract}
The classical (i.e. non-quantum) equilibrium statistical mechanics of a
two-dimensional one-component plasma (a system of charged
point-particles embedded in a neutralizing background) living on a
pseudosphere (an infinite surface of constant negative curvature) is
considered. In the case of a flat space, it is known that, for a
one-component plasma, there are several reasonable definitions of the 
pressure, and that some of them are not equivalent to each other. 
In the present paper, this problem is revisited in the case of a
pseudosphere. General relations between the different pressures are
given. At one special temperature, the model is exactly solvable in the
grand canonical ensemble. The grand potential and the one-body density 
are calculated in a disk, and the thermodynamic limit is investigated.
The general relations between the different pressures are checked on the
solvable model. 
\end{abstract}

\medskip


\medskip

\noindent LPT Orsay 02-67

\vfill

\noindent $^1$Dipartimento di Fisica Teorica dell' Universit\'a and
Istituto Nazionale di Fisica della Materia, Strada Costiera 11, 34014
Trieste, Italy;  
e-mail: rfantoni@ts.infn.it
 
\noindent $^2$Laboratoire de Physique Th\'eorique, B\^atiment 210, 
Universit\'e de Paris-Sud, 91405 Orsay, France (Unit\'e Mixte de 
Recherche no.8627-CNRS); e-mail: Bernard.Jancovici@th.u-psud.fr

\noindent $^3$Permanent address: Grupo de F\'{\i}sica T\'eorica de la
Materia Condensada, Departamento de F\'{\i}sica, Universidad de Los
Andes, A.A. 4976, Bogot\'a, Colombia; e-mail: gtellez@uniandes.edu.co

\newpage

\section{INTRODUCTION}
Coulomb systems such as plasmas or electrolytes are made of charged
particles interacting through Coulomb's law. The simplest model of a
Coulomb system is the one-component plasma (OCP), also called jellium:
an assembly of identical point charges, embedded in a neutralizing
uniform background of the opposite sign. Here we consider the classical
(i.e.~non-quantum) equilibrium statistical mechanics of the
OCP. Although many features of more realistic systems are correctly
reproduced, this model has the peculiarity that there are several
reasonable definitions of its pressure, and some of these definitions
are not equivalent to each other~\cite{Choquard,Navet}. 

The two-dimensional version (2D OCP) of the OCP has been much
studied. Provided that the Coulomb potential due to a point-charge is
defined as the solution of the Poisson equation in a two-dimensional
world (i.e.~is a logarithmic function $-\ln r$ of the distance $r$ to 
that point-charge), the 2D OCP mimicks many generic properties of the
three-dimensional Coulomb systems. Of course, this toy logarithmic model
does not describe real charged particles, such as electrons, confined on
a surface, which nevertheless interact through the three dimensional
Coulomb potential $1/r$. One motivation for studying the 2D OCP is that 
its equilibrium statistical mechanics is exactly solvable at one special
temperature: both the thermodynamical quantities and the correlation
functions are available~\cite{Janco1}.  

How the properties of a system are affected by the curvature of the
space in which the system lives is a question which arises in general
relativity. This is an incentive for studying simple models. Thus, the
problem of a 2D OCP on a pseudosphere has been considered~\cite
{JancoTellez}. A pseudosphere is a non-compact Riemannian surface of
constant negative curvature. Unlike the sphere it has an infinite area
and it is not embeddable in the three dimensional Euclidean space. The
property of having an infinite area makes it interesting from the point
of view of Statistical Physics because one can take the thermodynamic
limit on it. 

For this 2D OCP on a pseudosphere, the problem of studying and
comparing the different possible definitions of the pressure also
arises. This is the subject of the present paper. In
Section~\ref{sec:pseudo-OCP}, we give some basic properties of the
pseudosphere and of a 2D OCP on it. In Section~\ref{sec:pressures}, we
define the different pressures and derive general relations between
them. In Section~\ref{sec:Gamma=2}, we illustrate the general
properties by considering the special temperature at which all
properties can be explicitly and exactly calculated.

\section{PSEUDOSPHERE AND ONE-COMPONENT PLASMA} 
\label{sec:pseudo-OCP}

\subsection{The Pseudosphere}
There are at least three commonly known sets of coordinates to
describe a pseudosphere of Gaussian curvature $-1/a^2$. The 
one which renders explicit the resemblance with the sphere is $\vec{q}
=(q^1,q^2)=(q^\tau,q^\varphi)=(\tau,\varphi)$ with $\tau\in[0,\infty[$ 
and $\varphi\in[0,2\pi[$, the metric being 
\begin{equation}
ds^2=g_{\alpha\beta}\,dq^\alpha dq^\beta
=a^2(d\tau^2+\sinh^2\tau\, d\varphi^2)
\end{equation} \label{2.1}
Another set of coordinates often used is $(r,\varphi)$ with $r/(2a)
=\tanh(\tau/2)$. They are the polar coordinates of a disk of radius
$2a$. The metric in terms of these new coordinates is
\begin{equation} \label{2.2}
ds^2=\frac{dr^2+r^2d\varphi^2}{[1-(r^2/4a^2)]^2}
\end{equation}
The disk with such a metric is called the Poincar\'e disk. Through an
appropriate conformal transformation, the Poincar\'e disk can be mapped
onto the upper half-plane with some metric, the Poincar\'e 
half-plane, but this latter representation will not be used here. 
The geodesic distance $d_{01}$ between any two points
$\vec{q}_0=(\tau_0,\varphi_0)$ and $\vec{q}_1=(\tau_1,\varphi_1)$ on 
the pseudosphere is
given by
\begin{equation} \label{2.3}
\cosh(d_{01}/a)=\cosh\tau_1\cosh\tau_0-\sinh\tau_1\sinh\tau_0
\cos(\varphi_1-\varphi_0)
\end{equation}

Given the set of points at a geodesic distance from the origin less than
or equal to $d$, that we shall call a disk of radius $d$, we can easily
determine its circumference
\begin{equation} \label{2.4}
{\cal C}=2\pi\, a\, \sinh\left(\frac{d}{a}\right)
\begin{array}[t]{c}
\mbox{{\Huge$\sim$}}\\[-10pt]
\mbox{\scriptsize $d\rightarrow
\infty$}
\end{array}
\pi \, a\, e^{d/a}
\end{equation}
and its area
\begin{equation} \label{2.5}
{\cal A}=4\pi\,a^2\, \sinh^2\left(\frac{d}{2a}\right)
\begin{array}[t]{c}
\mbox{{\Huge$\sim$}}\\[-10pt]
\mbox{\scriptsize $d\rightarrow
\infty$}
\end{array}
\pi\,a^2\, e^{d/a}
\end{equation}
 
The Laplace-Beltrami operator on the pseudosphere is
\begin{equation} \label{2.6}
\Delta=\frac{1}{a^2}\left(\frac{1}{\sinh\tau}\frac{\partial}{\partial
\tau}\sinh\tau \frac{\partial}{\partial\tau}+\frac{1}{\sinh^2\tau}
\frac{\partial^2}{\partial\varphi^2}\right)
\end{equation}

\subsection{The One-Component Plasma}
The 2D OCP which is considered here is an ensemble of $N$ identical
point particles of charge $q$, constrained to move in a disk
of radius $d=a\tau_0$ by an infinite potential
barrier on the boundary of this domain $\tau=\tau_0$. The average
particle number density is $n=N/{\cal A}$, where ${\cal A}$ is the area
(\ref{2.5}). There is a background with a charge density $\rho_b=-qn_b$ 
uniformly smeared on the disk ($\rho_b$ is 0 outside the disk).
It is convenient to introduce the number of elementary charges in the
background: $N_b=n_b{\cal A}$. The total charge is not necessarily 0, 
thus in general $n_b\neq n$.

The pair Coulomb potential $v(d)$ between two unit charges, a geodesic 
distance $d$ apart, satisfies the Poisson equation on the pseudosphere, 
\begin{equation} \label{2.7}
\Delta v(d)=-2\pi\delta^{(2)}(d)
\end{equation}
where $\delta^{(2)}(d)$ is the Dirac delta function on the curved 
manifold. This Poisson equation admits a solution vanishing at infinity,
\begin{equation} \label{2.8}
v(d)=-\ln\left[\tanh\left(\frac{d}{2a}\right)\right]
\end{equation}

The electrostatic potential of the background $w(\vec{q})$ satisfies
\begin{equation} \label{2.9}
\Delta w(\vec{q})=-2\pi\rho_b
\end{equation}
By symmetry, this electrostatic potential is only a function of $\tau$.
Expressing the Laplacian (\ref{2.6}) in terms of the variable 
$\cosh \tau$, and requesting the solution to be regular at $\tau=0$ and
to have the correct value at $\tau=\tau_0$ (corresponding to the 
background total charge), one finds the solution

\begin{equation} \label{2.10}
w(\tau)=2\pi a^2qn_b\left\{\ln\left[\frac{1-\tanh^2(\tau_0/2)}
{1-\tanh^2(\tau/2)}\right]+\sinh^2(\tau_0/2)\ln[\tanh^2(\tau_0/2)]
\right\}
\end{equation}

Let $dS=2\pi a^2\,\sinh\tau\,d\tau$ be an area element. The self energy
of the background is 
\begin{eqnarray} \label{2.11}
v_0 &=& \frac{1}{2}\int_{\tau<\tau_0}\rho_bw(\tau)\,dS \nonumber \\
&=&(2\pi a^2qn_b)^2\{\sinh^2(\tau_0/2)-\ln[\cosh^2(\tau_0/2)]-
\sinh^4(\tau_0/2)\ln[\tanh^2(\tau_0/2)]\}
\end{eqnarray}

The total potential energy of the system is 
\begin{equation} \label{2.12}
U=v_0+v_{pb}+v_{pp}
\end{equation}
where $v_{pp}$ is the potential energy due to the interactions between
the particles,
\begin{equation} \label{2.13}
v_{pp}=\frac{1}{2}\sum_{\stackrel{i,j=1}{i\neq j}}^N q^2\,v(d_{ij})
\end{equation}
and $v_{pb}$ is the potential energy due to the interaction between
the particles and the background,
\begin{equation} \label{2.14} 
v_{pb}=\sum_{i=1}^N q\,w(\tau_i)
\end{equation}

\section{THE DIFFERENT PRESSURES AND THEIR RELATIONS} 
\label{sec:pressures}

In the case of a flat system, the pressure which is often considered,
termed the thermal pressure, is defined from the free energy $F$ by the
standard relation $P^{(\theta)}=-(\partial F/\partial {\cal A})_
{\beta,N,N_b}$, where $\beta$ is the inverse temperature. In the case of
a flat neutral $(N=N_b)$ 2D OCP, this thermal pressure is given by the
simple exact expression $\beta P^{(\theta)}=n[1-(\beta q^2/4)]$~\cite
{Salzberg,Hauge}. Thus, this thermal pressure becomes negative for 
$\beta q^2>4$, i.e.~at low temperatures. This pathology of the OCP
occurs also in three dimensions; it is related to the presence of an
inert background without kinetic energy. Indeed, the uniform background
can be considered as the limit of a gas of negative particles of charge
$-\epsilon$ and number density $\nu$, when $\epsilon\rightarrow 0$, 
$\nu\rightarrow\infty$, for a fixed value of the charge density
$-\epsilon\nu$. In this limit, the ideal-gas part (kinetic part) of the
background average energy density becomes infinite. In the OCP
Hamiltonian, this infinite energy density is omitted. The price paid for
this omission is that the corresponding (infinite) ideal-gas
contribution to the pressure is omitted, and the remaining pressure may
be negative.\setcounter{footnote}{3}\footnote{In the case of a
two-dimensional \emph{two}-component plasma made of point-particles, the
pressure also becomes negative when extrapolated to low temperatures 
$\beta q^2>4$. However, now $\beta q^2>4$ is outside the domain of
definition of the partition function.}  

Unhappy with this negativeness, Choquard et al.\cite{Choquard} and Navet
et al.\cite{Navet} have  introduced another pressure, the kinetic
pressure $P^{(k)}$, which is the pressure exerted on the wall by the
particles of charge $q$ only. This 
kinetic pressure turns out to be also the one which is obtained through
the use of the virial theorem. Although for usual fluids the thermal and 
kinetic pressures are equivalent, in the presence of a background they
are different, with the kinetic pressure being always positive. This
positiveness led the above authors to argue that the kinetic pressure is
the ``right'' one. Anyhow, a detailed comparison of the diverse possible
definitions of the pressure of a flat OCP has been done~\cite{Choquard}.

In the present paper, it is this comparison that we extend to the case
of a 2D OCP on a pseudosphere. We shall restrict ourselves to the case
of a domain in the shape of a disk. We are especially interested in the
thermodynamic limit, i.e.~when the disk radius becomes infinite, for
fixed values of $\beta,n,n_b$.

\subsection{Kinetic and Virial Pressures}
The average force exerted by the particles on a perimeter element $ds$
is $(1/\beta)n^{(1)}(\tau_0)ds$, where $n^{(1)}(\tau)$ is the one-body
density at the distance $a\tau$ from the origin. Therefore, the kinetic
pressure is  
\begin{equation} \label{3.1}
P^{(k)}=(1/\beta)n^{(1)}(\tau_0)
\end{equation}
We assume that this quantity has a limit when
$\tau_0\rightarrow\infty$.  In Section~\ref{sec:Gamma=2}, this
assumption will be checked in the special case $\beta q^2=2$. It will
now be shown that the virial pressure $P^{(v)}$, i.e.~the pressure
computed from the virial theorem, is the same as $P^{(k)}$.

In terms of the $2N$ coordinate components $q^N$ and $2N$ conjugate
momentum components $p^N$, the Hamiltonian of our OCP of $N$ particles
is
\begin{equation} \label{3.2}
H(q^N,p^N)=T(q^N,p^N)+{\bar U}(q^N)
\end{equation}
where ${\bar U}=U+$confining potential and the kinetic energy $T$ is
\begin{equation} \label{3.3}
T=\frac{1}{2m}\sum_{i=1}^Ng^{\alpha\beta}(\vec{q}_i)p_{i\alpha}p_{i\beta}
\end{equation}
The Roman indices label the particles, and the lower or upper Greek
indices denote covariant or contravariant components, respectively. As
usual, a sum over repeated Greek indices is tacitly assumed.
The equations of motion for particle $i$ are
\begin{equation} \label{3.4}
\left\{
\begin{array}{l}
\displaystyle \dot{q}_i^{\alpha}=\frac{\partial H}{\partial
p_{i\alpha}}=\frac{1}{m} g^{\alpha \beta}(\vec{q}_i)p_{i\beta}\\\\
\displaystyle \dot{p}_{i\alpha}=-\frac{\partial H}{\partial
q_i^{\alpha}}=-\frac{1}{2m} \frac{\partial g^{\beta\gamma}}{\partial
q_i^{\alpha}}p_{i\beta}p_{i\gamma}-\frac{\partial\bar{U}}
{\partial q_i^{\alpha}}
\end{array}
\right.
\end{equation}
where the dot stands for total derivative with respect to time.
If we take the time derivative of $\sum_i q_i^{\tau}p_{i\tau}=
\sum_i \tau_ip_{i\tau}$, we find
\footnote{One may be tempted to start with the time derivative of 
$\sum_i q_i^{\alpha}p_{i\alpha}=\sum_i(\tau_ip_{i\tau}+
\varphi_ip_{i\varphi})$. Note however that this quantity 
does not remain finite at all times. This is because, when one follows
the motion of a particle colliding with the boundary, it may go around 
the origin indefinitely, and $\varphi_i$ (which must be defined as a 
continuous variable, without any $2\pi$ jumps) may increase
indefinitely. Thus the time average of the time derivative of 
this quantity does not vanish.}
\begin{equation} \label{3.5}
\frac{d}{dt}\sum_i \tau_ip_{i\tau}=\frac{1}{m}\sum_{i=1}^N 
g^{\tau\beta}(\vec{q}_i)p_{i\tau}p_{i\beta}-\frac{1}{2m}
\sum_{i=1}^N \tau_i\frac{\partial g^{\beta\gamma}}{\partial \tau_i}
p_{i\beta}p_{i\gamma}-\sum_{i=1}^N \tau_i\frac{\partial\bar{U}}
{\partial\tau_i}
\end{equation}
where the last term is called the virial of the system.
Since the system is confined in a finite domain, the coordinates
$\tau_i(t)$ and their canonically conjugated momenta
$p_{i\tau}(t)$ remain finite at all times. Thus,
\begin{equation} \label{3.6}
\left\langle\frac{d}{dt}\sum_{i=1}^N \tau_ip_{i\tau}\right\rangle_t=0
\end{equation}
where $\langle\ldots\rangle_t$ denotes a time average. Assuming that the
system is ergodic, we can replace time averages by microcanonical
averages. Assuming the equivalence of ensembles in the thermodynamic
limit, we can as well use canonical or grand-canonical averages
$\langle\ldots\rangle$. In the present section, we use canonical 
averages. The average of the r.h.s. of (\ref{3.5})
vanishes. Separating in the last term of (\ref{3.5}) the contribution
from the forces exerted by the walls, which is, in the average, 
$-a\tau_0{\cal C}P^{(v)}$, we obtain
\begin{equation} \label{3.7}
a\tau_0{\cal C}P^{(v)}=\left\langle\frac{1}{m}\sum_{i=1}^N
g^{\tau\beta}(\vec{q}_i)p_{i\tau}p_{i\beta}\right\rangle-
\left\langle\frac{1}{2m} \sum_{i=1}^N \tau_i\frac{\partial
g^{\beta\gamma}}{\partial \tau_i}
p_{i\beta}p_{i\gamma}\right\rangle-\left\langle\sum_{i=1}^N
\tau_i\frac{\partial U} {\partial\tau_i}\right\rangle
\end{equation}

We now calculate the three terms in the r.h.s. of (\ref{3.7}). The first
one is the average of twice a contribution to the Hamiltonian, which is
quadratic in the $N$ variables $p_{i\tau}$ ($g$ is diagonal); since the
average of a quadratic term in the Hamiltonian is $1/(2\beta)$, the
first term in the r.h.s. of (\ref{3.7}) is 
\begin{equation} \label{3.8}
\left\langle\frac{1}{m}\sum_{i=1}^N
g^{\tau\tau}(\vec{q}_i)(p_{i\tau})^2\right\rangle=\frac{N}{\beta}
\end{equation}

The second term reduces to $-\left\langle(1/2m)\sum_{i=1}^N\tau_i
(\partial g^{\varphi\varphi}/\partial
\tau_i)(p_{i\varphi})^2\right\rangle$.  Averaging first on
$p_{i\varphi}$ replaces $(p_{i\varphi})^2/2m$ by $1/[2\beta
g^{\varphi\varphi}(\tau_i)]$. The second term becomes
\begin{equation} \label{3.9}
\frac{1}{\beta}\left\langle\sum_{i=1}^N\frac{\tau_i}{\tanh\tau_i}\right
\rangle=
\frac{1}{\beta}\int_{\tau<\tau_0}n^{(1)}(\tau)\frac{\tau}{\tanh\tau}dS
\end{equation}

Finally, since
\begin{equation} \label{3.10}
\frac{dn^{(1)}(\tau_1)}{d\tau_1}=-\beta N\frac{\int e^{-\beta U}
(\partial U/\partial \tau_1)dS_2\ldots dS_N}{\int e^{-\beta U}
dS_1dS_2\ldots dS_N}
\end{equation}
the third term can be written as
\begin{equation} \label{3.11}
-N\left\langle\tau_1\frac{\partial U}{\partial\tau_1}\right\rangle=
\frac{1}{\beta}\int_{\tau_1<\tau_0}\tau_1\frac{dn^{(1)}(\tau_1)}
{d\tau_1}dS_1
\end{equation}

Putting together the contributions (\ref{3.8}),(\ref{3.9}), and 
(\ref{3.11}) gives for (3.7)
\begin{equation} \label{3.12}
a\tau_0{\cal C}P^{(v)}=\frac{N}{\beta}+\frac{1}{\beta}
\int_0^{\tau_0}\left[n^{(1)}(\tau)\frac{\tau}{\tanh\tau}
+\tau\frac{dn^{(1)}(\tau)}{d\tau}\right]2\pi a^2\sinh\tau\,d\tau
\end{equation}
After an integration by parts, (\ref{3.12}) becomes
\begin{equation} \label{3.13}
P^{(v)}=\frac{1}{\beta}n^{(1)}(\tau_0)=P^{(k)}
\end{equation}

\subsection{The Thermal Pressure}
The thermal pressure is defined as
\begin{equation} \label{3.14}
P^{(\theta)}=-\left(\frac{\partial F}{\partial {\cal A}}\right)_
{\beta,N,N_b}
\end{equation}
where $F$ is the free energy. This expression is appropriate for the
canonical ensemble, since $F$ is related to the canonical partition
function $Z$ by $\beta F=-\ln Z$.

\subsubsection{The Thermal Pressure in the Grand Canonical ensemble}
In the following, we shall also need an expression of the thermal
pressure appropriate for the grand canonical ensemble. It should be
remembered that, for a flat OCP in three dimensions, the grand canonical 
partition function must be defined~\cite{Lieb} as an ensemble of systems 
with any number $N$ of particles in a fixed volume and {\it with a fixed
background charge density} $-qn_b$ (using an ensemble of neutral
systems, i.e.~varying $n_b$ together with $N$ would give a divergent
grand partition function). In two dimensions, $\beta$ times
the free energy for a neutral flat system ~\cite{Janco1} behaves as
$[1-(\beta q^2/4)]N\ln N$ as $N\rightarrow\infty$, and therefore
the neutral grand canonical partition function diverges if 
$\beta q^2>4$.
This indicates that, in the present case of a 2D OCP on a pseudosphere, 
a similar divergence might occur for an ensemble of neutral systems, and
we prefer to use an ensemble  with a fixed background (which,
furthermore, will be seen to be exactly solvable at $\beta q^2=2$). 
Thus, the grand partition function $\Xi$ and the corresponding grand
potential $\Omega=-(1/\beta)\ln\Xi$ are functions of $\beta,{\cal
A},\zeta,n_b$, where $\zeta$ is the fugacity. The usual Legendre
transformation from $F$ to $\Omega$ and from $N$ to $\zeta$ changes
(\ref{3.14}) into
\begin{equation} \label{3.15}
P^{(\theta)}=-\left(\frac{\partial\Omega}{\partial {\cal A}}\right)_
{\beta,\zeta,N_b}
\end{equation}
We assume that, even on a pseudosphere, the grand potential is
extensive, i.e.~of the form $\Omega={\cal A}\omega(\beta,\zeta,n_b)$.
Since $\omega$ depends on ${\cal A}$ through $n_b=N_b/{\cal A}$, 
equation~(\ref{3.15}) becomes
\begin{equation} \label{3.16}
P^{(\theta)}=-\omega+n_b\frac{\partial\omega}{\partial n_b}
\end{equation}
Note the difference with an ordinary fluid, without a background, for
which $P^{(\theta)}=-\omega$.

\subsubsection{The $P^{(\theta)}-P^{(k)}$ Difference}
\label{sec:ptheta-pk}
For a OCP, the thermal pressure is different from the kinetic pressure. 
In the case of a 2D OCP in a flat disk, in the thermodynamic limit, the
boundary becomes a straight line and the difference was found to
be~\cite{Choquard}
\begin{equation} \label{3.17}
P^{(\theta)}-P^{(k)}=-2\pi q^2n_b\int_0^{\infty}[n^{(1)}(x)-n_b]x\,dx
\end{equation} 
where $ n^{(1)}(x)$ is the density at distance $x$ from the boundary.
Using the Poisson equation, one can write (\ref{3.17}) in the equivalent
form~\cite{Totsuji}
\begin{equation} \label{3.18}
P^{(\theta)}-P^{(k)}=qn_b[\phi_{\rm surface}-\phi_{\rm bulk}]
\end{equation}
where $\phi_{\rm bulk}$ and $\phi_{\rm surface}$ are the electric
potential in the bulk and on the disk boundary,
res\-pec\-ti\-ve\-ly.\footnote{In the original 
papers~\cite{Choquard,Totsuji}, the derivations of (\ref{3.17}) and 
(\ref{3.18}) have been done in the case of a neutral system. However,
these derivations can be easily extended to systems carrying a total non
vanishing charge.} 
 
Equation~(\ref{3.18}) can be proven as follows. Either in the flat case,
or in the case of a pseudosphere, let us consider a large disk of area 
${\cal A}$, filled with a 2D OCP. For compressing it
infinitesimally, changing the area by $d{\cal A}<0$, at constant
$\beta,N,N_b$, we must provide the reversible work $\delta W
=-P^{(\theta)}d{\cal A}$. We may achieve that compression in two
steps. First, one compresses the particles only, leaving the background
behind; the corresponding work is $\delta W^{(1)} =-P^{(k)}d{\cal A}$,
since $P^{(k)}$ is the force per unit length exerted on the wall by the
particles alone. Then, one compresses the background, i.e.~brings the 
charge $qn_b\,d{\cal A}$ from a region where the potential is 
$\phi_{\rm surface}$ into the plasma where the potential is $\phi(r)$,
spreading it uniformly; the corresponding work is $\delta W^{(2)}=
qn_b\,d{\cal A}[(1/{\cal A})\int \phi(r) dS-\phi_{\rm surface}]$, where
$\phi(r)$ is the potential at distance $r$ from the center. Therefore,
\begin{equation} \label{3.19}
P^{(\theta)}-P^{(k)}=
qn_b[\phi_{\rm surface}-\frac{1}{{\cal A}}\int\phi(r)\,dS]
\end{equation}
We expect $\phi(r)$ to differ from $\phi_{\rm bulk}$ only in the neighborhood of 
the boundary circle. 

In the case of a flat disk, the contribution of
this neighborhood to the integral in (\ref{3.19}) is negligible in the
thermodynamic limit, $\phi(r)$ can be replaced by the constant
$\phi_{\rm bulk}$, and one obtains (\ref{3.18}). On a pseudosphere, 
(\ref{3.19}) [with $\phi(\tau)$ instead of $\phi(r)$] is still
valid. However, now, in the large-disk limit, the integration element 
$dS=2\pi a^2\sinh\tau\,d\tau$ makes the boundary neighborhood dominant,
and we rather have 
\begin{equation} \label{3.20}
P^{(\theta)}-P^{(k)}\sim qn_b\left[\phi(\tau_0)-
e^{-\tau_0}\int_0^{\tau_0}\phi(\tau)e^\tau\,d\tau\right]
\end{equation}
After some manipulations, in the thermodynamic limit, (\ref{3.20}) can
be shown to be equivalent to   
\begin{equation} \label{3.21}
P^{(\theta)}-P^{(k)}=-2\pi a^2 n_bq^2\int_0^{\infty}[n^{(1)}(\sigma)
-n_b]\,\sigma e^{-\sigma}\,d\sigma
\end{equation}
where we have introduced the variable $\sigma=\tau_0-\tau$ and 
$n^{(1)}(\sigma)$ now denotes the particle density at distance $a\sigma$
from the boundary. Indeed, in (\ref{3.21}), $n^{(1)}(\tau)-n_b$ can
be expressed in terms of $\phi(\tau)$ through the Poisson equation 
$\Delta \phi(\tau)=-2\pi q[n^{(1)}(\tau)-n_b]$. Since the charge
density is localized at large $\tau$, we can use for the Laplacian 
$\Delta \sim a^{-2}[d^2/d\tau^2 + d/d\tau]$. After integrations by
parts, (\ref{3.20}) is recovered.

In conclusion, (\ref{3.17}) valid for a large flat disc generalizes 
into (\ref{3.21}) on a pseudosphere. In the limit $a\rightarrow\infty,\,
\sigma\rightarrow 0,\,a\sigma=x$, equation~(\ref{3.21}) does reproduce
(\ref{3.17}).

\subsection{The Mechanical Pressure}

Choquard et al.\cite{Choquard} have also defined a mechanical pressure,
in terms of the free energy $F$, as
\begin{equation} \label{3.22}
P^{(m)}=-\left(\frac{\partial F}{\partial {\cal A}}\right)_{\beta,N,n_b}
\end{equation}
In terms of the grand potential $\Omega$, a Legendre transformation now
gives
\begin{equation} \label{3.23}
P^{(m)}=-\left(\frac{\partial \Omega}{\partial {\cal A}}\right)
_{\beta,\zeta,n_b}
\end{equation}
If the grand potential is extensive, of the form $\Omega=
{\cal A}\omega(\beta,\zeta,n_b)$, (\ref{3.23}) gives
\begin{equation} \label{3.24}
P^{(m)}=-\omega
\end{equation}

The difference $P^{(m)}-P^{(k)}$ can be obtained by a slight change in
the argument of Section~\ref{sec:ptheta-pk}. Again, we consider a
large disk filled with a 2D OCP of area ${\cal A}$, and we compress it
infinitesimally, changing its area by $d{\cal A}<0$, now at constant
$\beta,N,n_b$, providing the reversible work $\delta W=-P^{(m)}d{\cal
A}$, in two steps. Again, first one compresses the particles only,
leaving the background behind, and the corresponding work is $\delta
W^{(1)} =-P^{(k)}d{\cal A}$. Then, one must withdraw the leftover
background charge $qn_b\,d{\cal A}$, bringing it from the surface
where the potential is $\phi_{\rm surface}$ to infinity where the
potential vanishes. The corresponding work is $\delta W^{(2)}=
-qn_b\,d{\cal A}\,\phi_{\rm surface}$. Therefore, for a disk on a
pseudosphere, $P^{(m)}-P^{(k)}=qn_b\phi_{\rm surface}$.\footnote{ This
result is identical with the one obtained by Choquard et
al.\cite{Choquard} in the case of a flat disk. However, their general
formula might make difficulties in two dimensions, because the Coulomb
potential $-\ln(r/L)$ does not vanish at infinity and involves an
arbitrary constant length $L$. These difficulties do not arise on a
pseudosphere.}

In the thermodynamic limit, 
$\phi_{\rm surface}\rightarrow 2\pi a^2q(n-n_b)$ and
\begin{equation} \label{3.25}
P^{(m)}-P^{(k)}=2\pi a^2q^2n_b(n-n_b)
\end{equation}
This difference vanishes for a neutral system ($n=n_b$).  

The relations~(\ref{3.21}) and~(\ref{3.25}) between the different
pressures obtained here by means of electrostatic arguments can also
be obtained in a more formal way following Choquard et
al.\cite{Choquard}, using the dilatation method (doing a change of
variable $\tau=\tau_0 \tilde{\tau}$ in the partition function to
explicitly show the area ${\cal A}$ dependence) and the BGY equations
to replace the two-body density terms that appear in the calculations
by one-body density terms.

\subsection{The Maxwell Tensor Pressure}

On a pseudosphere, since the area of a large domain is of the same order
as the area of the neighborhood of the boundary, all the above
definitions of the pressure depend on the boundary conditions. In
previous papers, a definition of a bulk pressure independent of the 
boundary conditions has been looked for. After an erroneous 
attempt~\cite{JancoTellez}, it has been argued~\cite{Janco2,Janco3} that
a bulk pressure $P_{\rm Maxwell}$ could be defined from the Maxwell
stress tensor~\cite{Jackson} at some point well inside the fluid. The
result was 
\begin{equation} \label{3.26}
\beta P_{\rm Maxwell}=n_b\left(1-\frac{\beta q^2}{4}\right)
\end{equation}
That same equation of state holds for the 2D OCP on a plane, a sphere,
or a pseudosphere.

\section{\uppercase{Exact results at} $\beta q^2=2$}
\label{sec:Gamma=2}

When the Coulombic coupling constant is $\beta q^2=2$, all the
thermodynamic properties and correlation functions of the
two-dimensional one-component plasma can be computed exactly in
several geometries~\cite{Janco1,JancoAlastueyOCP,CaillolSphere}
including the pseudosphere~\cite{JancoTellez}. In
reference~\cite{JancoTellez} the density and correlation functions in
the bulk, on a pseudosphere, were computed. Here we are interested in
these quantities near the boundary. In reference~\cite{JancoTellez}
the calculations were done for a system with a $-\ln\sinh(d/2a)$
interaction and it was shown that this interaction gives the same
results as the real Coulomb interaction $-\ln\tanh(d/2a)$, as far as the
bulk properties are concerned. The argument in favor of this equivalence
no longer holds for the density and other quantities near the boundary; 
therefore we shall concentrate on the real Coulomb system with a 
$-\ln\tanh(d/2a)$ interaction. This system was briefly considered in the
Appendix of reference~\cite{JancoTellez}. For the sake of completeness,
we revisit here the reduction of the statistical mechanics problem to 
the study of a certain operator.

\subsection{The grand potential}

Working with the set of coordinates $(r,\varphi)$ on the pseudosphere
(the Poincar\'e disk representation), the particle $i$-particle $j$
interaction term in the Hamiltonian can be written as~\cite{JancoTellez}
\begin{equation} \label{4.1}
v(d_{ij})=-\ln\tanh(d_{ij}/2a)=
-\ln\left|
\frac{(z_i-z_j)/(2a)}{1-(z_i\bar{z}_j/4a^2)}
\right|
\end{equation}
where $z_j=r_j e^{i\varphi_j}$ and $\bar{z}_j$ is the complex conjugate
of $z_j$. This interaction (\ref{4.1}) happens to be the Coulomb 
interaction in a flat disc of radius $2a$ with ideal conductor walls.
Therefore, it is possible to use the techniques which have been 
developed~\cite{Forrester,JancoTellez1} for dealing with ideal conductor 
walls, in the grand canonical ensemble.

The grand canonical partition function of the OCP at
fugacity $\zeta$ with a fixed background density $n_b$, when $\beta
q^2=2$, is 
\begin{equation}
\Xi=C_0 \left[1+\sum_{N=1}^{\infty} \frac{1}{N!}
\int \prod_{k=1}^N \frac{\zeta(r_k)r_k dr_k d\varphi_k}
{[1-(r_k^2/4a^2)]}
\prod_{i<j}\left|\frac{(z_i-z_j)/(2a)}{1-(z_i
\bar{z}_j/4a^2)}\right|^2 \right]
\end{equation}
where for $N=1$ the product $\prod_{i<j}$ must be replaced by 1. 
We have defined a position-dependent fugacity $\zeta(r)=\zeta
[1-r^2/(4a^2)]^{4\pi n_b a^2-1} e^C$ which includes the
particle-background interaction~(\ref{2.10}) and only one factor
$[1-r^2/(4a^2)]^{-1}$ from the integration measure
$dS=[1-r^2/(4a^2)]^{-2}\,d\r$. This should prove to be convenient
later. The $e^C$ factor is
\begin{equation}
\label{eq:cste-exp-c}
e^C=\exp\left[4\pi n_b a^2\left(\ln\cosh^2\frac{\tau_0}{2}
-\sinh^2\frac{\tau_0}{2}\ln\tanh^2\frac{\tau_0}{2}\right)\right]
\end{equation}
which is a constant term coming from the particle-background
interaction term~(\ref{2.10}) and
\begin{equation}
\label{eq:logC0}
\ln C_0=\frac{(4\pi n_b a^2)^2}{2}\left[
\ln\cosh^2\frac{\tau_0}{2}+\sinh^2\frac{\tau_0}{2}\,\left(
\sinh^2\frac{\tau_0}{2}
\ln\tanh^2\frac{\tau_0}{2}-1
\right)
\right]
\end{equation}
which comes from the background-background
interaction~(\ref{2.11}). Notice that for large domains, when
$\tau_0\to\infty$, we have
\begin{equation}
\label{eq:cste-exp-c-asymptot}
e^C \sim \left[ \frac{e^{\tau_0+1}}{4} \right]^{4\pi n_b a^2}
\end{equation}
and
\begin{equation}
\label{eq:logC0-asymptot}
\ln C_0 \sim -\frac{(4\pi n_b a^2)^2 e^{\tau_0}}{4}
\end{equation}
Let us define a set of reduced complex coordinates $u_i=(z_i/2a)$
inside the Poincar\'e disk and its corresponding images
$u_i^*=(2a/\bar{z}_i)$ outside the disk. By using Cauchy identity
\begin{equation}
\det
\left(
\frac{1}{u_i-u_j^*}
\right)_{(i,j)\in\{1,\cdots,N\}^2}
=
(-1)^{N(N-1)/2}\:
\frac{\prod_{i<j} (u_i-u_j)(u^*_i-u^*_j)}{\prod_{i,j} (u_i-u_j^*)}
\end{equation}
the particle-particle interaction term together with the
$[1-(r_i^2/4a^2)]^{-1}$ other term from the integration measure can be
cast into the form
\begin{equation}
\prod_{i<j}\left|
\frac{(z_i-z_j)/(2a)}{1-(z_i \bar{z}_j/4a^2)}
\right|^2
\prod_{i=1}^N [1-(r_i^2/4a^2)]^{-1}
=
\det
\left(
\frac{1}{1-u_i \bar{u}_j}
\right)_{(i,j)\in\{1,\cdots,N\}^2}
\end{equation}
The grand canonical partition function then is
\begin{equation}
\label{eq:GrandPart-prelim}
\Xi= \left[1+\sum_{N=1}^{\infty} \frac{1}{N!}
\int \prod_{k=1}^N d^2\r_k \zeta(r_k)
\det
\left(
\frac{1}{1-u_i \bar{u}_j}
\right)
\right]
C_0
\end{equation}

We shall now show that this expression can be reduced to an infinite
continuous determinant, by using a functional integral representation
similar to the one which has been developed for the two-component
Coulomb gas~\cite{ZinnJustin}. Let us consider the Gaussian partition 
function 
\begin{equation}
\label{eq:part-fun-Grass-libre}
Z_0=\int {\cal D}\psi {\cal D}\bar{\psi} \,\exp\left[\int \bar{\psi}(\r)
M^{-1}(z,\bar{z}') \psi(\r')\, d^2\r\, d^2\r'\right]
\end{equation}
The fields $\psi$ and $\bar{\psi}$ are anticommuting Grassmann
variables.  The Gaussian measure in~(\ref{eq:part-fun-Grass-libre}) is
chosen such that its covariance is equal to\footnote{Actually the
operator $M$ should be restricted to act only on analytical functions
for its inverse $M^{-1}$ to exist.}
\begin{equation}
\left<\bar{\psi}(\r_i)\psi(\r_j)\right>
=
M(z_i,\bar{z}_j)=\frac{1}{1-u_i \bar{u}_j}
\end{equation}
where $\langle\ldots\rangle$ denotes an average taken with the Gaussian
weight of (\ref{eq:part-fun-Grass-libre}). By construction we have
\begin{equation} \label{Z_0}
Z_0=\det(M^{-1})
\end{equation}
Let us now consider the following partition function
\begin{equation}
Z=\int {\cal D}\psi {\cal D}\bar{\psi} \exp\left[\int \bar{\psi}(\r)
M^{-1}(z,\bar{z}') \psi(\r') d^2\r d^2\r' +\int \zeta(r)
\bar{\psi}(\r)\psi(\r) \,d^2\r \right]
\end{equation}
which is equal to
\begin{equation}
Z=\det(M^{-1}+\zeta)
\end{equation}
and then
\begin{equation} \label{Z/Z_0}
\frac{Z}{Z_0}=\det[M(M^{-1}+\zeta)]=\det[1+K]
\end{equation}
where 
\begin{equation} \label{K}
K(\r,\r')=M(z,\bar{z}')\,\zeta(r')=
\frac{\zeta(r')}{1-u\bar{u}'}
\end{equation}
The results which follow can also be obtained by exchanging the order of
the factors $M$ and $M^{-1}+\zeta$ in (\ref{Z/Z_0}), i.e. by replacing 
$\zeta(r')$ by $\zeta(r)$ in (\ref{K}), however using the definition 
(\ref{K}) of $K$ is more convenient. Expanding the ratio $Z/Z_0$ in 
powers of $\zeta$ we have
\begin{equation}
\label{eq:expans-ZZ0}
\frac{Z}{Z_0}=
1+
\sum_{N=1}^{\infty}
\frac{1}{N!}
\int \prod_{k=1}^N d^2\r_k
\zeta(r_k)
\left<\bar{\psi}(\r_1)\psi(\r_1)\cdots
\bar{\psi}(\r_N)\psi(\r_N)\right>
\end{equation}
Now, using Wick theorem for anticommuting variables~\cite{ZinnJustin},
we find that
\begin{equation}
\label{eq:WickFerms}
\left<\bar{\psi}(\r_1)\psi(\r_1)\cdots
\bar{\psi}(\r_N)\psi(\r_N)\right>
=\det M(z_i,\bar{z}_j)=\det\left(\frac{1}{1-u_i \bar{u}_j}\right)
\end{equation}
Comparing equations~(\ref{eq:expans-ZZ0})
and~(\ref{eq:GrandPart-prelim}) with the help of
equation~(\ref{eq:WickFerms}) we conclude that\footnote{Actually, the
determinants $Z_0$ and $Z$ are divergent quantities, since the
eigenvalues of M (restricted to act on analytical functions) are 
easily found to be $4\pi a^2/(\ell +1)$, with $\ell$ any non-negative
integer. However, the ratio $Z/Z_0$ turns out to be finite.}
\begin{equation}
\label{eq:Xi-det}
\Xi=C_0\,\frac{Z}{Z_0}=C_0\det(1+K)
\end{equation}

The problem of computing the grand canonical partition function has
been reduced to finding the eigenvalues of the operator $K$. The
eigenvalue problem for $K$ reads
\begin{equation}
\label{eq:vpK}
\int \zeta e^C 
\frac{\left(\displaystyle 1-\frac{r'^2}{4a^2}\right)^{4\pi n_b a^2-1}}%
{\displaystyle 1-\frac{z\bar{z}'}{4a^2}}\,
\Phi(\r')\,r'\,dr'd\varphi'
=
\lambda \Phi(\r)
\end{equation}
For $\lambda\neq 0$ we notice from equation~(\ref{eq:vpK}) that
$\Phi(\r)=\Phi(z)$ is an analytical function of $z$. Because of
the circular symmetry it is natural to try
$\Phi(z)=\Phi_{\ell}(z)=z^{\ell}=r^{\ell}e^{i\ell\varphi}$ with $\ell$
a non-negative integer (the functions $z^{\ell}$ form a complete basis
for the analytical functions). Expanding
\begin{equation}
\frac{1}{\displaystyle1-\frac{z\bar{z}'}{4a^2}}=
\sum_{n=0}^{\infty}
\left(\frac{z\bar{z}'}{4a^2}\right)^{n}
\end{equation}
and replacing $\Phi_{\ell}(z)=z^{\ell}$ in equation~(\ref{eq:vpK}) one
can show that $\Phi_{\ell}$ is actually an eigenfunction of $K$ with
eigenvalue
\begin{equation}
\label{eq:lambda-vp-de-K}
\lambda_{\ell}=
4\pi a^2 \zeta e^C
B_{t_0}(\ell+1,4\pi n_b a^2)
\end{equation}
with $t_0=r_0^2/(4a^2)=\tanh^2 (\tau_0/2)$ and 
\begin{equation}
B_{t_0}(\ell+1,4\pi n_b a^2)=
\int_0^{t_0} (1-t)^{4\pi n_b a^2-1} t^{\ell} \, dt
\end{equation}
the incomplete beta function. So we finally arrive to the result
for the grand potential 
\begin{equation}
\label{eq:grand-potential-somme}
\beta\Omega = -\ln\Xi 
=
-\ln C_0
-
\sum_{\ell=0}^{\infty}
\ln\left(
1+4\pi a^2 \zeta e^C B_{t_0}(\ell+1,4\pi n_b a^2)
\right)
\end{equation}
with $e^C$ and $\ln C_0$ given by equations~(\ref{eq:cste-exp-c})
and~(\ref{eq:logC0}). This result is valid for any disk domain of
radius $a\tau_0$. Later, in Section~\ref{sec:large-domains}, we will
derive a more explicit expression of the grand potential for large
domains $\tau_0\to\infty$.

\subsection{The density}
As usual one can compute the density by doing a functional derivative
of the grand potential with respect to the position-dependent fugacity:
\begin{equation}
\label{eq:n-funct-deriv}
n^{(1)}(\r)=
\left(1-\frac{r^2}{4a^2}\right)^{2}
\zeta(r)\frac{\delta\ln\Xi}{\delta \zeta(r)}
\end{equation}
The factor $[1-(r^2/4a^2)]^2$ is due to the
curvature~\cite{JancoTellez}, so that $n^{(1)}(\r)\, dS$ is the average
number of particles in the surface element $dS=[1-(r^2/4a^2)]^{-2}\,d\r$.
Using a Dirac-like notation, one can formally write
\begin{equation}
\ln\Xi=\Tr \ln(1+K)+\ln C_0=
\int \left<\r\left|
\ln(1+\zeta(r)M)\right|\r\right>
\,d\r
+\ln C_0
\end{equation}
Then, doing the functional derivative~(\ref{eq:n-funct-deriv}), one
obtains
\begin{equation}
n^{(1)}(\r)= \left(1-\frac{r^2}{4a^2}\right)^{2}
\zeta(r)\left<\r\left| (1+K)^{-1}M \right|\r\right> =
4\pi a\left(1-\frac{r^2}{4a^2}\right)^2\zeta(r)\tilde{G}(\r,\r)
\end{equation}
where we have defined $\tilde{G}(\r,\r')$ by\footnote{the factor $4\pi
a$ is there just to keep the same notations as in 
reference~\cite{JancoTellez}.} $\tilde{G}=(1+K)^{-1}M/(4\pi a)$. More
explicitly, $\tilde{G}$ is the solution of $(1+K)\tilde{G}=M/(4\pi a)$, 
that is
\begin{equation}
\label{eq:eq-Green-function}
\tilde{G}(\r,\r') + \zeta e^C \int
\tilde{G}(\r'',\r')\,\frac{\left(\displaystyle
1-\frac{r''^2}{4a^2}\right)^{4\pi n_b a^2-1}}{\displaystyle
1-\frac{z\bar{z}''}{4a^2}}\, d\r'' = \frac{1}{\displaystyle 4\pi a
\left[1-\frac{z\bar{z}'}{4a^2}\right]}
\end{equation}
and the density is given by
\begin{equation}
n^{(1)}(\r)=4\pi a \zeta e^C \left(1-\frac{r^2}{4a^2}\right)^{4\pi n_b
a^2+1}
\tilde{G}(\r,\r) 
\end{equation}
 From the integral equation~(\ref{eq:eq-Green-function}) one can see
that $\tilde{G}(\r,\r')$ is an analytical function of $z$. Thus the
solution is of the form
\begin{equation}
\tilde{G}(\r,\r')=\sum_{\ell=0}^{\infty} a_{\ell}(\r') z^{\ell}
\end{equation}
and equation~(\ref{eq:eq-Green-function}) yields
\begin{equation}
\label{eq:solution-G}
\tilde{G}(\r,\r')=
\frac{1}{4\pi a}
\sum_{\ell=0}^{\infty}
\left(\frac{z\bar{z}'}{4a^2}\right)^{\ell}
\frac{1}{1+4\pi a^2 \zeta e^{C} B_{t_0} (\ell+1,4\pi n_b a^2)}
\end{equation}
Then the density is given by
\begin{equation}
\label{eq:densite-somme}
n^{(1)}(r)=\zeta e^C \left(1-\frac{r^2}{4a^2}\right)^{4\pi n_b a^2+1}
\sum_{\ell=0}^{\infty}
\left(\frac{r^2}{4a^2}\right)^{\ell}
\frac{1}{1+4\pi a^2 \zeta e^{C} B_{t_0} (\ell+1,4\pi n_b a^2)}
\end{equation}
After some calculation (see the Appendix), it can be shown that, in the
limit $a\rightarrow\infty$, the result for the flat disk in the
canonical ensemble \cite{jancoch} 
\begin{equation} \label{surf}
\frac{n^{(1)}(r)}{n_b}=\exp(-\pi n_b r^2)
\sum_{\ell =0}^{N_b-1}\frac{(\pi n_b r^2)^{\ell}}{\gamma
(\ell +1,\,N_b)}
\end{equation}
is recovered, up to a correction due to the non-equivalence of ensembles
in finite systems. In (\ref{surf}), $\gamma$ is the incomplete gamma
function
\begin{equation} \label{gamma}
\gamma(\ell +1,\,x)=\int_0^x t^{\ell}e^{-t}dt
\end{equation} 
In that flat-disk case, in the thermodynamic limit (half-space), 
$n^{(1)}(r_0)=n_{\mathrm{contact}}\rightarrow n_b\ln\,2$.

\subsection{Large domains}
\label{sec:large-domains}
We are now interested in large domains $\tau_0\to\infty$. In this
thermodynamic limit we will show that the sums in
equations~(\ref{eq:grand-potential-somme})
and~(\ref{eq:densite-somme}) can be replaced by integrals. For
pedagogical reasons we will first consider the case $4\pi n_b a^2=1$ in
which the calculations are simpler, and afterwards deal with the general
case.

\subsubsection{The case $4\pi n_b a^2=1$}
In this case the incomplete beta function that appears in
equations~(\ref{eq:grand-potential-somme})
and~(\ref{eq:densite-somme}) simply is
\begin{equation}
B_{t_0}(\ell+1, 1)=
\frac{t_0^{\ell+1}}{\ell+1}=\frac{[\tanh^2 (\tau_0/2)]^{\ell+1}}{\ell+1}
\end{equation}
When $\tau_0\to\infty$ we have
\begin{equation}
B_{t_0}(\ell+1, 1)\sim
\frac{\exp(-4(\ell+1)e^{-\tau_0})}{\ell+1}
\end{equation}
Then the sum appearing in the grand
potential~(\ref{eq:grand-potential-somme}) takes the form
\begin{equation}
\label{eq:somme-de-Riemann}
\sum_{\ell=0}^{\infty}
\ln\left(
1+\frac{\zeta
e}{n_b}\frac{\exp(-4(\ell+1)e^{-\tau_0})}{4(\ell+1)e^{-\tau_0}}
\right)
\end{equation}
where we have used the asymptotic
expression~(\ref{eq:cste-exp-c-asymptot}) for $e^C$. This sum can be
seen as a Riemann sum for the variable
$x=4(\ell+1)e^{-\tau_0}$. Indeed, for large values of $\tau_0$, the
variable $x$ varies in small steps $dx=4e^{-\tau_0}$. The
sum~(\ref{eq:somme-de-Riemann}) then converges, when $\tau_0\to\infty$, to
the integral
\begin{equation}
\int_0^{\infty}
\ln \left(1+\frac{\zeta e}{n_b}\frac{e^{-x}}{x}\right)\,
\frac{dx}{4e^{-\tau_0}}
\end{equation}
This expression together with equation~(\ref{eq:logC0-asymptot}) for
the constant $\ln C_0$ gives the grand potential in the thermodynamic
limit $\tau_0\to\infty$
\begin{equation}
\beta\Omega\sim -\frac{e^{\tau_0}}{4}
\left[
\int_0^{\infty}
\ln\left(1+\frac{\zeta e}{n_b}\frac{e^{-x}}{x}\right)
\,dx - 1
\right]
\end{equation}
We notice that the grand potential is extensive as expected. The area
of the system being ${\cal A}=4\pi a^2 \sinh^2(\tau_0/2)\simeq \pi a^2
e^{\tau_0}$, we find that the grand potential per unit area
$\omega=\Omega/{\cal A}$ is given by
\begin{equation}
\beta \omega
=
-n_b
\left[
\int_0^{\infty}
\ln\left(1+\frac{\zeta e}{n_b}\frac{e^{-x}}{x}\right)
\,dx - 1
\right]
\end{equation}
Similar calculations lead from equation~(\ref{eq:densite-somme}) to
the density $n^{(1)}(\sigma)$ near the boundary as a function of the
distance from that boundary $a\sigma=a(\tau_0-\tau)$,
\begin{equation}
n^{(1)}(\sigma)= \zeta e\, e^{2\sigma} \int_0^{\infty} \frac{e^{-x
e^{\sigma}}}{\displaystyle 1+\frac{\zeta e}{n_b}\frac{e^{-x}}{x}}
\,dx
\end{equation}
After the change of variable $xe^{\sigma}\to x$, this can be written as
\begin{equation}
\label{eq:densite-profile-alpha=1}
\frac{n^{(1)}(\sigma)}{n_b}=
\int_0^{\infty}
\frac{x e^{-x}\,dx}{\displaystyle \frac{x e^{-\sigma}}{(\zeta e/n_b)}+
e^{-xe^{-\sigma}}}
\end{equation}
The average density $n=N/{\cal A}$ can be obtained integrating the
density profile~(\ref{eq:densite-profile-alpha=1}) or by using the
thermodynamic relation $N=-\beta\zeta(\partial \Omega/\partial
\zeta)$. We find 
\begin{equation}
\frac{n}{n_b}=
\int_0^{\infty} \frac{e^{-x}\,dx}{\displaystyle
\frac{x}{(\zeta e/n_b)}+e^{-x}}
\end{equation}

\subsubsection{The general case}
With the case $4\pi n_b a^2=1$ we have illustrated the general
procedure for computing the thermodynamic limit. Now we proceed to
compute it in the more general case where $4\pi n_b a^2$ has any
positive value. To simplify the notations let us define $\alpha=4\pi
n_b a^2$. The main difficulty is to find a suitable asymptotic
expression of the incomplete beta function 
\begin{equation}
B_{t_0}(\ell+1,\alpha)
=
\int_0^{t_0}
(1-t)^{\alpha-1} t^{\ell}\,dt
\end{equation}
when $t_0\to 1$ which is valid for large $\ell$. As we have noticed in the
previous section the main contribution from the sum in $\ell$ that
appears in the grand potential comes from large values of $\ell$ which
are of order $e^{\tau_0}$. For these values of $\ell $ the integrand in
the definition of the beta function $(1-t)^{\alpha-1} t^{\ell}$ is very
peaked around $t=t_0$ and decays very fast when $t\to 0$. So the main
contribution to the incomplete beta function comes from values of $t$
near $t_0$. It is then natural to do the change of variable in the
integral $t=t_0-v$ where with the new variable $v$ the integral is
mainly dominated by small values of $v$. Then we have
\begin{equation}
B_{t_0}(\ell+1,\alpha)=
\int_0^{t_0} (1-t_0+v)^{\alpha-1} e^{\ell\ln(t_0-v)}
\,dv
\end{equation}
Replacing $t_0$ by its asymptotic value $t_0\sim 1-4e^{-\tau_0}$ and
taking into account  that $v$ is small (of order $e^{-\tau_0}$), we
find, at first order in $e^{-\tau_0}$,
\begin{equation}
B_{t_0}(\ell+1,\alpha)\sim
\frac{1}{\ell^{\alpha}}\,
\Gamma(\alpha,x)
\end{equation}
where we have introduced once more the variable $x=4\ell e^{-\tau_0}$
(at first order in $e^{-\tau_0}$ it is the same variable $x=4(\ell+1)
e^{-\tau_0}$ introduced in the case $\alpha=1$) and
\begin{equation}
\Gamma(\alpha,x)
=
\int_x^{\infty}
y^{\alpha-1}e^{-y}\,dy
\end{equation}
is an incomplete gamma function. With this result and
equation~(\ref{eq:cste-exp-c-asymptot}) the term $e^C
B_{t_0}(\ell+1,\alpha)$ in the
expressions~(\ref{eq:grand-potential-somme})
and~(\ref{eq:densite-somme}) appears as a function of the continuous
variable $x=4\ell e^{-\tau_0}$
\begin{equation}
e^C
B_{t_0}(\ell+1,\alpha)
\sim
e^{\alpha}\,
\frac{\Gamma(\alpha,x)}{x^{\alpha}}
\end{equation}
With this result we can replace the sums for $\ell$ in
equations~(\ref{eq:grand-potential-somme})
and~(\ref{eq:densite-somme}) by integrals over the variable $x$ and we
find the following expressions for the grand potential per unit area
\begin{equation}
\label{eq:resultat-grand-pot}
\beta\omega=
\frac{1}{4\pi a^2}
\left\{
(4\pi n_b a^2)^2
-
\int_0^\infty
\ln\left[
1+4\pi a^2\zeta e^{4\pi n_b a^2} \frac{\Gamma(4\pi n_b a^2,x)}{x^{4\pi
n_b a^2}}
\right]
\, dx
\right\}
\end{equation}
and the density
\begin{equation}
\label{eq:resultat-densite-1}
n^{(1)}(\sigma)
=
\zeta e^{4\pi n_b a^2} e^{(4\pi n_b a^2+1)\sigma}
\int_0^{\infty}
\frac{e^{-xe^{\sigma}}\ dx}{\displaystyle
1+ 4\pi a^2 \zeta e^{4\pi n_b a^2}\frac{\Gamma(4\pi n_b a^2,x)}{x^{4\pi
n_b a^2}}}
\end{equation}
In particular the contact value of the density, that is when $\sigma=0$, is
\begin{equation}
\label{eq:densite-contact}
n_{\mathrm{contact}}=n^{(1)}(0)
=
\zeta e^{4\pi n_b a^2} 
\int_0^{\infty}
\frac{e^{-x }\ dx}{\displaystyle
1+ 4\pi a^2 \zeta e^{4\pi n_b a^2}\frac{\Gamma(4\pi n_b a^2,x)}{x^{4\pi
n_b a^2}}}
\end{equation}
After some calculation (see the Appendix), it can be shown that, in the
limit $a\rightarrow\infty$, the result for the flat disk in the
thermodynamic limit $n_{\mathrm{contact}}=n_b\ln\,2$ is again recovered.

An alternative expression for the density which we will also use is
obtained by doing the change of variable $xe^{\sigma}\to x $ and
introducing again $\alpha=4\pi n_b a^2$
\begin{equation}
\label{eq:resultat-densite-2}
\frac{n^{(1)}(\sigma)}{n_b}
=
\int_0^{\infty}
\frac{x^{\alpha} e^{-x}\ dx}{\displaystyle
\frac{x^{\alpha} 
e^{-\alpha \sigma}}{(\zeta e^{\alpha}/n_b)}
+
\alpha\Gamma(\alpha,xe^{-\sigma})}
\end{equation}
 From this expression it can be seen that in the bulk, when
$\sigma\to\infty$ and $e^{-\sigma}\to 0$, the density is equal to the
background density, $n^{(1)}(\sigma)\to n_b$. The system is neutral in
the bulk. The excess charge, which is controlled by the fugacity
$\zeta$, concentrates as usual on the boundary.

The average total number of particles $N$ and the average density
$n=N/{\cal A}$ can be computed either by using the thermodynamic
relation 
\begin{equation}
N=-\beta\zeta\frac{\partial \Omega}{\partial \zeta}
\end{equation}
or by integrating the density profile~(\ref{eq:resultat-densite-1})
\begin{equation}
N=\int_{\tau<\tau_0} 
n^{(1)}(\sigma)\, dS
=\pi a^2
e^{\tau_0}\int_0^{\infty} n^{(1)}(\sigma)\, e^{-\sigma}\,d\sigma
\end{equation}
The two methods yield the same result, as expected,
\begin{equation}
\label{eq:densite-moyenne-1}
n=\frac{N}{{\cal A}}=
\zeta e^{4\pi n_b a^2}
\int_0^{\infty}
\frac{\Gamma(4\pi n_b a^2,x)\,dx}{\displaystyle
x^{4\pi n_b a^2}+4\pi a^2 \zeta e^{4\pi n_b a^2}\Gamma(4\pi n_b
a^2,x)}
\end{equation}
The ratio of the average density and the background density can
be put in the form
\begin{equation}
\label{eq:densite-moyenne-2}
\frac{n}{n_b}=
\int_0^{\infty}
\frac{\Gamma(\alpha,x)\, dx}{\displaystyle
\frac{x^{\alpha}}{(\zeta e^{\alpha}/n_b)}+\alpha \Gamma(\alpha,x)}
\end{equation}
As seen on equations~(\ref{eq:resultat-densite-2})
and~(\ref{eq:densite-moyenne-2}) the density profile $n^{(1)}(\sigma)$
and the average density $n$ are functions of the parameter $g=\zeta
e^{4\pi n_b a^2}/n_b$.  Different values of this parameter $g$ give
different density profiles and mean densities. Depending on the value of
$g$ the system can be globally positive, negative or neutral. From
equation~(\ref{eq:densite-moyenne-2}) it can be seen that the average
density is a monotonous increasing function of the fugacity, as it
should be. Therefore there is one unique value of the fugacity for
which the system is globally neutral. For the case $4\pi n_b a^2=1$,
we have determined numerically the value of $g$ needed for the system
to be neutral, $n=n_b$. This value is $g=\zeta e/n_b=1.80237$.

It may be noted that, in the case of a flat disk in the grand canonical
ensemble, the 2D OCP remains essentially neutral (the modulus of its
total charge cannot exceed one 
elementary charge $q$), whatever the fugacity $\zeta$ might
be~\cite{Janco4,Janco5}; this is because the Coulomb interaction 
$-\ln (r/L)$ becomes infinite at infinity and bringing an excess charge 
from a reservoir at infinity to the system already carrying a net charge
would cost an infinite energy. On the contrary, in the present case of a
2D OCP on a pseudosphere, the Coulomb interaction (\ref{2.8}) has an
exponential decay at large distances, and varying the fugacity does
change the total charge of the disk.
  
Figure~\ref{fig:density} shows several plots of the density
$n^{(1)}(\sigma)$ as a function of the distance $\sigma$ from the 
boundary (in units of $a$), for different values of $g$, in the case 
$\alpha=4\pi n_b a^2=1$. It is interesting to notice that for $g\leq 1$ 
the density is always an increasing function of $\sigma$. Far away from
the boundary, the density approaches the background density 
$n_b$ from below. On the other hand when $g>1$, but not too large, the
density profile shows an oscillation: $n^{(1)}(\sigma)$ is no longer a
monotonous function of $\sigma$. Far away from the boundary,
$\sigma\to\infty$, the density now approaches the background density
from above. Finally, when $g$ is large enough, the density profile is
again monotonous, now a decreasing function of $\sigma$. 

The change of behavior as $\sigma\rightarrow\infty$ can actually be
shown analytically. Let us define
$u=e^{-\sigma}$. From equation~(\ref{eq:resultat-densite-2}) we have
\begin{equation}
\frac{\partial}{\partial u}\left( \frac{n^{(1)}(\sigma)}{n_b} \right)
=\int_0^{\infty} 
\frac{ \alpha x^{2\alpha} u^{\alpha-1} e^{-x-xu}}{\displaystyle
\left(\frac{(xu)^{\alpha}}{g}+\alpha \Gamma(\alpha,xu)\right)^2}
\left[1-\frac{e^{xu}}{g}\right]\,dx
\end{equation}
The first term in the integral is always positive. The second term,
$1-(e^{xu}/g)$, in the limit $\sigma\to\infty$ ($u\to0$) is
$1-(1/g)$. If $g<1$ it is negative, then $\partial n^{(1)}/\partial u$
is negative and $n^{(1)}(\sigma)$ is then an increasing function of
$\sigma$ when $\sigma\to\infty$ as it was noticed in the last
paragraph.

Also, in this case $\alpha=1$, when $\zeta\to\infty$ the density
profile~(\ref{eq:densite-profile-alpha=1}) can be computed explicitly
\begin{equation}
\frac{n^{(1)}(\sigma)}{n_b}=\frac{1}{\left(1-e^{-\sigma}\right)^2}
\end{equation}
It is clearly a monotonous decreasing function of $\sigma$.

\subsection{Relations between the different pressures}

 From the explicit expressions~(\ref{eq:resultat-grand-pot})
and~(\ref{eq:resultat-densite-2}) for the grand potential and the
density profile, we can check the relations between the different
pressures obtained in Section~\ref{sec:pressures}. The mechanical
pressure simply is $P^{(m)}=-\omega$ and it is given by
equation~(\ref{eq:resultat-grand-pot}). This expression can be
transformed by doing an integration by parts in the integral giving
\begin{equation}
\label{eq:p-meca-parcial}
\beta P^{(m)}=
-\frac{1}{4\pi a^2}
\left\{
\int_0^{\infty}
\frac{\displaystyle 4\pi a^2 x \zeta e^{4\pi n_b a^2} \frac{d}{dx}
\left[
\frac{\Gamma(4\pi n_b a^2,x)}{x^{4\pi n_b a^2}}\right]
}{\displaystyle
1+4\pi a^2 \zeta e^{4\pi n_b a^2} \frac{\Gamma(4\pi n_b
a^2,x)}{x^{4\pi n_b a^2}}}
\, dx+
(4\pi n_b a^2)^2
\right\}
\end{equation}
By the replacement
\begin{equation}
\frac{d}{dx}
\left[
\frac{\Gamma(4\pi n_b a^2,x)}{x^{4\pi n_b a^2}}\right]
=
-\frac{e^{-x}}{x}-4\pi n_b a^2 \frac{\Gamma(4\pi n_b a^2,x)}{x^{4\pi
n_b a^2 +1}}
\end{equation}
in equation~(\ref{eq:p-meca-parcial}), one recognizes the
expressions~(\ref{eq:densite-contact})
and~(\ref{eq:densite-moyenne-1}) for the contact density and the
average density, thus giving
\begin{equation}
\beta P^{(m)}=
n^{(1)}(0)+4\pi n_b a^2 (n - n_b)
\end{equation}
which is precisely, when $\beta q^2=2$, the relation~(\ref{3.25})
between the mechanical pressure $P^{(m)}$ and the kinetic pressure
$P^{(k)}=(1/\beta)n^{(1)}(0)$ obtained in Section~\ref{sec:pressures}.

The thermal pressure is 
\begin{equation}
P^{(\theta)}=-\omega(\zeta,n_b) + n_b \left(\frac{\partial
\omega(\zeta,n_b)}{\partial n_b}\right)_{\zeta}
\end{equation}
The last term in this equation is given by
\begin{equation}
\label{eq:partial-pres-term}
\beta n_b \frac{\partial \omega}{\partial n_b}=
\frac{1}{4\pi a^2}
\left\{
2\alpha^2 - \int_0^{\infty}
\frac{4\pi a^2 \zeta}{\displaystyle 
1+\frac{4\pi a^2 \zeta e^{\alpha}\Gamma(\alpha,x)}{x^{\alpha}}}
\,\alpha\frac{\partial}{\partial \alpha}
\left[
\frac{e^{\alpha}\Gamma(\alpha,x)}{x^{\alpha}}
\right]
\,dx
\right\}
\end{equation}
Making the replacement
\begin{equation}
\alpha\frac{\partial}{\partial \alpha}
\left[
\frac{e^{\alpha}\Gamma(\alpha,x)}{x^{\alpha}}
\right]
=
\alpha e^{\alpha}
\left(
\frac{\Gamma(\alpha,x)}{x^{\alpha}}
+
\frac{\partial}{\partial \alpha}
\left[\frac{\Gamma(\alpha,x)}{x^{\alpha}}\right]
\right)
\end{equation}
in equation~(\ref{eq:partial-pres-term}), one recognizes in the
first term the average density $n$, thus obtaining
\begin{equation}
\beta n_b\frac{\partial\omega}{\partial n_b}=
\alpha(2n_b-n)-\alpha I
\end{equation}
where 
\begin{equation}
I=
\int_0^{\infty}
\frac{\zeta e^{\alpha}}{\displaystyle 
1+\frac{4\pi a^2 \zeta e^{\alpha}\Gamma(\alpha,x)}{x^{\alpha}}}
\,\frac{\partial}{\partial \alpha}
\left[
\frac{\Gamma(\alpha,x)}{x^{\alpha}}
\right]
\,dx
\end{equation}
So the thermal pressure is given by
\begin{equation}
\beta P^{(\theta)}=
n^{(1)}(0)+\alpha n_b - \alpha I
\end{equation}
On the other hand the integral appearing in the general
relation~(\ref{3.21}) between the thermal pressure and the
kinetic pressure
\begin{equation}
J=
\int_0^{\infty} (n^{(1)}(\sigma)-n_b)\,e^{-\sigma}\sigma\,d\sigma
\end{equation}
can be split into two parts
\begin{equation}
J=-n_b + I'
\end{equation}
with
\begin{equation}
I'=\int_0^{\infty} n^{(1)}(\sigma)\sigma e^{-\sigma}\,d\sigma
\end{equation}
Using the actual integral representation for the density profile given
by equation~(\ref{eq:resultat-densite-1}) yields
\begin{equation}
I'=
\int_0^{\infty}
\frac{\zeta e^{\alpha}}{\displaystyle 
1+\frac{4\pi a^2 \zeta e^{\alpha}\Gamma(\alpha,x)}{x^{\alpha}}}
\left\{
\int_0^{\infty}
e^{\alpha\sigma} e^{-xe^{\sigma}} \sigma\,d\sigma
\right\}\,dx
\end{equation}
The integral over $\sigma$ can be cast in the form
\begin{equation}
\frac{\partial}{\partial\alpha}\left[
\int_0^{\infty} e^{\alpha\sigma} e^{-xe^{\sigma}}\,d\sigma
\right]
\end{equation}
By doing the change of variable $y=xe^{\sigma}$ one immediately
recognizes the integral representation of the incomplete gamma
function. The above expression is then equal to
\begin{equation}
\frac{\partial}{\partial\alpha}\left[
\frac{\Gamma(\alpha,x)}{x^{\alpha}}
\right]
\end{equation}
Thus we have proven that $I'=I$ and finally we have the relation
\begin{equation}
\beta (P^{(\theta)}-P^{(k)})=
-4\pi n_b a^2
\int_0^{\infty}
(n^{(1)}(\sigma)-n_b)\,e^{-\sigma}\sigma\,d\sigma
\end{equation}
which is relation~(\ref{3.21}) in the solvable case $\beta q^2=2$.

\section{\uppercase{Conclusion}}
In a flat space, the neighborhood of the boundary of a large domain has
a volume which is a negligible fraction of the whole volume. This is
why, for the statistical mechanics of ordinary fluids, usually there is
a thermodynamic limit: when the volume becomes infinite, quantities such
as the free energy per unit volume or the pressure have a unique limit,
independent of the domain shape and of the boundary conditions. However,
even in a flat space, the one-component plasma is special. For the OCP, 
it is possible to define several non-equivalent pressures, some of
which, for instance the kinetic pressure, obviously are
surface-dependent even in the infinite-system limit.

Even for ordinary fluids, statistical mechanics on a pseudosphere is
expected to have special features, which are essentially related to the 
property that, for a large domain, the area of the neighborhood of the 
boundary is of the same order of magnitude as the whole area. Although 
some bulk properties, such as correlation functions far away from the 
boundary, will exist, extensive quantities such as the free energy 
or the grand potential are strongly dependent on the boundary 
neighborhood and surface effects. For instance, in the large-domain
limit, no unique limit is expected for the free energy per unit area
$F/{\cal A}$ or the pressure 
$-(\partial F/\partial {\cal A})_{\beta,N}$.

In the present paper, we have studied the 2D OCP on a pseudosphere, for
which surface effects are expected to be important for both reasons:
because we are dealing with a one-component plasma and because the space
is a pseudosphere. Therefore, although the
correlation functions far away from the boundary have unique
thermodynamic limits~\cite{JancoTellez}, many other properties are 
expected to depend on the domain shape and on the boundary conditions.
This is why we have considered a special well-defined geometry: the
domain is a disk bounded by a plain hard wall, and we have studied the
corresponding large-disk limit. Our results have been derived only for 
that geometry.

We have been especially interested by different pressures which can be
defined for this system. It has been shown that the virial pressure 
$P^{(v)}$ (defined through the virial theorem) and the kinetic pressure 
$P^{(k)}$ (the force per unit length that the particles alone exert on
the wall) are equal to each other. We have also considered the thermal
pressure $P^{(\theta)}$, the definition of which includes contributions
from the background. It should be noted that this thermal pressure 
is also dependent on surface effects, since it is defined by 
(\ref{3.14}) and (\ref{3.15}) in terms of the free energy or the 
grand potential, and the corresponding partition functions
include relevant contributions from the surface region. The thermal
pressure is not equal to the previous ones. We have also considered 
the so-called mechanical pressure $P^{(m)}$ which differs from the 
kinetic one only for charged systems. General relations among these
different pressures have been established.

One of the referees of the present paper has asked which one of these
different pressures is the ``right'' one, i.e. which one would be
measured by a barometer. The answer, based on the previous paragraph, is
that it depends on which kind of ``barometer'' is used. For instance,
the measured pressure would not be the same if the barometer, placed on
the wall, measures only the force exerted on it by the particles alone,
or if it also feels the force exerted by the background.

When $\beta q^2=2$, the model is exactly solvable, in the grand canonical
ensemble. Explicit expressions have been obtained for the grand
potential, the density profile, and the  pressures. The general
relations between the different pressures have been checked.

A bulk pressure, independent of the surface effects, can be defined from
the Maxwell stress tensor. It is not astonishing that this bulk
pressure is different from the previous ones, all of which depend on
surface effects.

\section*{\uppercase{Appendix: the flat limit}}
\renewcommand{\theequation}{A.\arabic{equation}}

In this Appendix we study the flat limit $a\to\infty$ of the
expressions found for the density in section~\ref{sec:Gamma=2}.  We
shall study the limit $a\to\infty$ for a finite system and then take
the thermodynamic limit and compare to the result of taking first the
thermodynamic limit and then the flat limit $a\to\infty$. Since for a
large system on the pseudosphere boundary effects are of the same
order as bulk effects it is not clear a priori whether computing these
two limits in different order would give the same results. We shall
show that, indeed, the same results are obtained.

For a finite disk of radius $d=a\tau_0$, we have in the flat limit
$a\to\infty$, $d\sim r_0$. In equation~(\ref{eq:densite-somme}), in the
limit $a\to\infty$, the term $e^C$ given by (\ref{eq:cste-exp-c})
becomes
\begin{equation}
e^C\sim \left(\frac{r_0^2}{4a^2}\right)^{-N_b}
e^{N_b}
\end{equation}
where $N_b=\pi n_b r_0^2$ is the number of particles in the background
in the flat limit. Since for large $a$, $t_0=r_0^2/(4a^2)$ is small, the
incomplete beta function in equation~(\ref{eq:densite-somme}) is
\begin{equation}
B_{t_0}(\ell+1,\alpha)
=
\int_0^{t_0}
e^{(\alpha-1)\ln(1-t)}
\, t^{\ell}\,dt
\sim
\int_0^{t_0}
e^{-(\alpha-1)t}
\, t^{\ell}\,dt
\sim
\frac{\gamma(\ell+1,N_b)}{\alpha^{\ell+1}}
\end{equation}
Expanding $(1-(r^2/4a^2))^{4\pi n_b a^2}\sim \exp(-\pi n_b r^2)$ in
equation~(\ref{eq:densite-somme}) we finally find the density as a
function of the distance $r$ from the center
\begin{equation}
n^{(1)}(r)=
n_b e^{-\pi n_b r^2}
\sum_{\ell=0}^{\infty}
\frac{(\pi n_b r^2)^{\ell}}{\alpha^{\ell-N_b} N_b^{N_b} e^{-N_b}
(n_b/\zeta)+
\gamma(\ell+1,N_b)}
\end{equation}
When $\alpha\to\infty$ the terms for $\ell>N_b$ in the sum vanish
because $\alpha^{\ell-N_b}\to\infty$. Then
\begin{equation}
\label{eq:densite-flat}
n^{(1)}(r)=
n_b e^{-\pi n_b r^2}
\sum_{\ell=0}^{E(N_b)-1}
\frac{(\pi n_b r^2)^{\ell}}{\gamma(\ell+1,N_b)}
+\Delta n^{(1)}(r)
\end{equation}
The first term is the density for a flat OCP in the canonical ensemble
with a background with $E(N_b)$ elementary charges ($E(N_b)$ is the
integer part of $N_b$). The second term is a correction due to the
inequivalence of the ensembles for finite systems and it depends on
whether $N_b$ is an integer or not. If $N_b$ is not an integer
\begin{equation}
\Delta n^{(1)}(r)=n_b
\frac{(\pi n_b r^{2})^{E(N_b)} e^{-\pi n_b r^2}}{\gamma(E(N_b)+1,N_b)}
\end{equation}
and if $N_b$ is an integer
\begin{equation}
\Delta n^{(1)}(r)=n_b
\frac{(\pi n_b r^{2})^{N_b} e^{-\pi n_b r^2}}{N_b^{N_b} e^{-N_b}
(n_b/\zeta)+\gamma(N_b+1,N_b)}
\end{equation}
In any case in the thermodynamic limit $r_0\to\infty$,
$N_b\to\infty$, this term $\Delta n^{(1)}(r)$ vanishes giving the
known results for the OCP in a flat space in the canonical
ensemble~\cite{Janco1,jancoch}. Integrating the profile
density~(\ref{eq:densite-flat}) one finds the average number of
particles. For a finite system it is interesting to notice that the
average total number of particles $N$ is
\begin{equation}
N=E(N_b)+1
\end{equation}
for $N_b$ not an integer and
\begin{equation}
N=N_b+\frac{1}{\displaystyle 1+
\frac{N_b^{N_b}e^{-N_b} n_b}{\zeta \gamma(N_b+1,N_b)}}
\end{equation}
for $N_b$ an integer. In both cases the departure from the neutral
case $N=N_b$ is at most of one elementary charge as it was noticed
before~\cite{Janco4,Janco5}.

Let us now consider the other order of the limits. We start with the
expression~(\ref{eq:densite-contact}) for the contact density in the
thermodynamic limit in the pseudosphere and show that in the limit
$a\to\infty$ the value of the contact density reduces to the known
expression for a neutral OCP in a flat space at a hard
wall~\cite{jancoch}. We also show that in that limit the average
density is independent of the fugacity and equal to the background
density $n=n_b$.

Equation~(\ref{eq:densite-contact}) can be rewritten as
\begin{equation} \label{contact}
\frac{n_{\mathrm{contact}}}{n_b}=
\int_0^{\infty}\frac{x^{\alpha}e^{-x}\,dx}{\frac{n_b}{\zeta}
x^{\alpha}e^{-\alpha} + \alpha\Gamma(\alpha,x)}
\end{equation}
For large $\alpha$, the numerator of the integrand in (\ref{contact})
has a sharp peak at $x=\alpha$ and can be expanded as
\begin{equation} \label{col}
x^{\alpha}e^{-x}\sim e^{\alpha\ln\alpha -\alpha -
\left(\frac{x-\alpha}{\sqrt{2\alpha}}\right)^2}
\end{equation}
In the denominator, using the large $\alpha$ expansion of the 
incomplete gamma function~\cite{Erdelyi}, and neglecting 1 with respect
to $\alpha$, we obtain 
\begin{equation} \label{alphaGamma}
\alpha\Gamma(\alpha,x)\sim \alpha^{\alpha} e^{-\alpha}
\sqrt{\frac{\pi\alpha}{2}}
\left[1-\erf\left(\frac{x-\alpha +1}{\sqrt{2\alpha}}\right)\right]
\end{equation}
where
\begin{equation}
\erf(t)=\frac{2}{\sqrt{\pi}}
\int_0^t e^{-u^2}\,du
\end{equation}
is the error function. Using (\ref{col}) and (\ref{alphaGamma}) 
in (\ref{contact}) gives
\begin{equation}
\frac{n_{\mathrm{contact}}}{n_b}\sim\int_0^{\infty}\frac
{e^{-\left(\frac{x-\alpha}{\sqrt{2\alpha}}\right)^2}dx}
{\frac{n_b}{\zeta}\left(\frac{x}{\alpha}\right)^{\alpha}
+\sqrt{\frac{\pi\alpha}{2}}
\left[1-\erf\left(\frac{x-\alpha +1}{\sqrt{2\alpha}}\right)\right]}
\end{equation}
For $x>\alpha$, the first term in the denominator goes to infinity for
large $\alpha$ and the integrand goes to zero. 
On the other hand, when $x<\alpha$, this same first term goes to zero, 
thus, after the change of variable $t=(x-\alpha)/\sqrt{2\alpha}$,
\begin{equation}
\frac{n_{\mathrm{contact}}}{n_b}\sim \frac{2}{\sqrt{\pi}}
\int_{-\sqrt{\alpha/2}}^0\frac{e^{-t^2}\,dt}{1-
\erf\left(t+\frac{1}{\sqrt{2\alpha}}\right)}
\end{equation}
Finally, as $\alpha\rightarrow\infty$,
\begin{equation}
\frac{n_{\mathrm{contact}}}{n_b}\rightarrow\int_{-\infty}^0
\frac{\frac{d\,\erf(t)}{dt}}{1-\erf(t)}dt=\ln\,2
\end{equation}
This is the known value~\cite{jancoch} for the contact density
at a hard plain wall for a neutral OCP.

Following the same lines, equation~(\ref{eq:densite-moyenne-2}) for
the average density becomes in the limit $\alpha\to\infty$
\begin{equation}
\frac{n}{n_b}
\sim
\sqrt{\frac{2}{\alpha}}\int_{-\sqrt{\alpha/2}}^{0}
\frac{\left[1-\erf(t)\right]\,dt}{1-\erf(t)} =1
\end{equation}
The average density is equal to the background density and it is
independent of the fugacity. Whatever value the fugacity has, the system
cannot be charged in the flat case in the thermodynamic limit.

\section*{\uppercase{Acknowledgments}}
R.~F.~would like to thank F.~Benatti for his lectures in Trieste and
stimulating discussions. B.~J.~and G.~T.~acknowledge support from 
ECOS Nord/COLCIENCIAS-ICFES-ICETEX action C00P02 of French and 
Colombian cooperation. G.~T.~acknowledge partial financial support 
from COLCIENCIAS and BID through project \#1204-05-10078.

%%%%%%%%%%%%%%%%%%%%%%%%%%%%%%%%
%
% List of figure captions
%
\newpage
\subsection*{List of figure captions}
\makeatletter
\def\@captype{figure}
\makeatother

%%%%%%%%%%%%%%%%%%%%%%%%%%%%%%%
%
%   Figure 1: density
%
\caption{
\label{fig:density}
The density profile $n^{(1)}(\sigma)$ (in units of $n_b$) as a
function of the distance from the boundary $\sigma$ (in units of $a$)
for different
values of the parameter $g=\zeta e/n_b$ in the case $4\pi n_b
a^2=1$. From bottom to top, in full line $g=0.5, 1.5, 2.5, 5.0, 10.0$
and in dashed line $g=1$ (change of behavior between monotonous
increasing profile and oscillating profile), $g=1.80237$ (globally
neutral system) and $g\to\infty$.
}

%%%%%%%%%%%%%%%%%%%%%%%%%%%%%
%
%   Figure 1: density
%
\newpage
\hbox{}
\vfill
\begin{center}
\includegraphics[width=10cm]{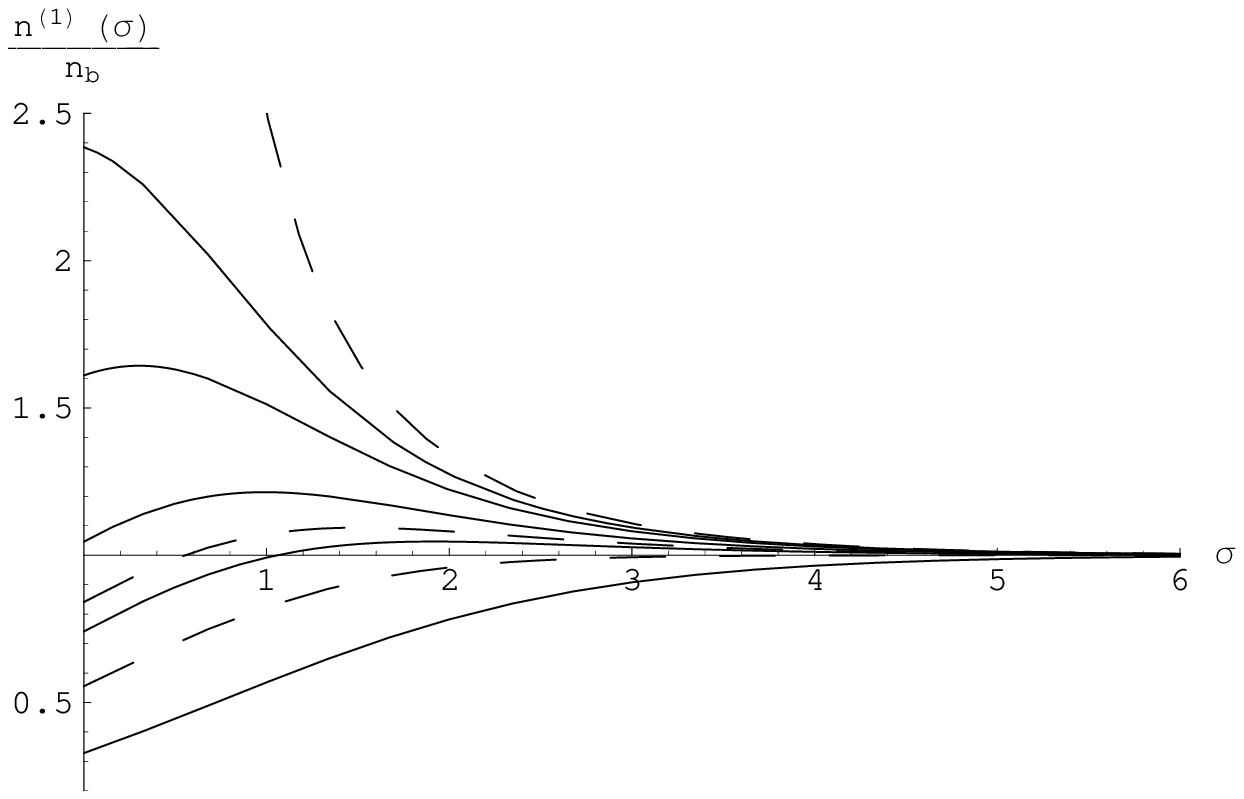}
\end{center}
\vfill
\begin{center}
R.~Fantoni, B.~Jancovici and G.~T\'ellez\hskip 5cm Figure 1
\end{center}
\thispagestyle{empty}
\end{document}